\documentclass[twocolumn, superscriptaddress, prb, preprintnumbers, showpacs, floatfix]{revtex4-1}

\usepackage{ifthen} 
\usepackage{graphicx}
\usepackage{amsmath}
\usepackage{braket}
\usepackage{cancel}
\usepackage{hyperref}
\hypersetup{backref,pdfpagemode=FullScreen,colorlinks=true,allcolors=blue}
\usepackage{color}

\def\edit#1{\textcolor{black}{#1}}

\begin{document}

\title{\edit{Inapplicability of exact constraints and 
a minimal two-parameter generalization  
to the DFT+$U$ based correction of 
self-interaction error}}

\author{Glenn Moynihan}
\email{omuinneg@tcd.ie}
\affiliation{School of Physics, CRANN and AMBER, 
Trinity College Dublin, Dublin 2, Ireland}

\author{Gilberto Teobaldi}
\affiliation{Stephenson Institute for Renewable Energy and  Department of Chemistry, The University of Liverpool, L69 3BX Liverpool, United Kingdom}
\affiliation{Beijing Computational Science Research Center, Beijing 100094, China}

\author{David D. O'Regan}
\affiliation{School of Physics, CRANN and AMBER, 
Trinity College Dublin, Dublin 2, Ireland}

\date{\today{}}

\begin{abstract}
In approximate density functional theory (DFT), 
the  self-interaction error is \edit{an electron delocalization anomaly}
associated with underestimated insulating gaps.
It exhibits a predominantly quadratic energy-density curve
that is amenable to correction using
 efficient, constraint-resembling 
methods such as DFT + Hubbard $U$ (DFT+$U$).
Constrained DFT (cDFT) enforces conditions 
on DFT \edit{exactly}, by means of 
self-consistently optimized Lagrange multipliers,
and \edit{while} its use to automate \edit{error} 
corrections is a compelling possibility,
we show that \edit{it is limited by a fundamental
incompatibility with} constraints 
beyond linear order.
\edit{We circumvent this problem by utilizing separate
linear and quadratic correction terms,
which may be interpreted either as distinct constraints, 
each with its own Hubbard $U$ type Lagrange multiplier, 
or as the  components of a generalized DFT+$U$ functional.
The latter approach prevails in our tests on a model one-electron 
system, $H_2^+$,
in that it readily recovers the exact total-energy
while symmetry-preserving pure constraints fail to do so.
%
%
%
The generalized DFT+$U$ functional moreover enables the simultaneous
correction of the total-energy and ionization potential, 
or the correction of either together with the 
enforcement of Koopmans' condition. 
%
For the latter case, 
we outline a practical, approximate scheme by which the 
required pair of Hubbard parameters, denoted 
as $U_1$ and $U_2$, 
may be calculated from first-principles.}

\end{abstract}


\maketitle

Approximate density-functional theory 
(DFT)~\cite{PhysRev.136.B864, PhysRev.140.A1133}
underlies much of contemporary 
quantum-mechanical atomistic simulation, providing
a widespread and valuable complement to 
experiment~\cite{RevModPhys.87.897,jain2016computational}.
The predictive capacity of DFT is
severely limited, however, by 
systematic errors~\cite{PhysRevLett.49.1691,Cohen792,PhysRevB.82.115121} 
exhibited by 
tractable exchange-correlation
functionals such as the local-density
approximation (LDA)~\cite{PhysRevB.23.5048} or 
generalized-gradient
approximations~\cite{PhysRevLett.77.3865}. 
Perhaps the most prominent of these pathologies
is the delocalization or many-electron 
self-interaction  error 
(SIE)~\cite{PhysRevB.23.5048}, 
which is due to spuriously curved 
rather than correctly piecewise-linear total-energy
profiles with respect to the total electron 
number~\cite{Cohen792,PhysRevLett.49.1691}. 
This  gives rise to the underestimated 
fundamental  gaps~\cite{PhysRevLett.49.1691},
charge-transfer energies~\cite{doi10.1021/cr200148b}, 
and reaction barriers~\cite{PhysRevLett.97.103001}
characteristic of practical DFT. 
\edit{While the construction of viable, 
explicit density functionals 
free of pathologies such as SIE 
is extremely challenging~\cite{doi:10.1021/cr200107z},
significant progress has  been made in 
the development of implicit functionals in the guise
of corrective approaches. 
Examples include  methods
that operate by correcting SIE on a one-electron basis
according to variationally optimized definitions, such as   
generalized~\cite{PhysRevLett.65.1148,
:/content/aip/journal/jcp/140/12/10.1063/1.4869581,PhysRevB.93.165120} Perdew-Zunger~\cite{PhysRevB.23.5048} approaches, 
in which much progress has recently been made
by generalizing to complex-valued
orbitals~\cite{PhysRevA.85.062514,
PhysRevLett.108.146401,
:/content/aip/journal/jcp/137/12/10.1063/1.4752229}, 
and those which  address many-electron SIE directly, such as
Koopman's compliant functionals~\cite{PhysRevB.82.115121,
PhysRevB.90.075135}.}
%
%
%

%
An established, computationally very efficient 
DFT correction scheme is 
DFT+$U$~\cite{PhysRevB.43.7570,PhysRevB.44.943,
PhysRevB.48.16929,PhysRevB.57.1505,
PhysRevB.58.1201,PhysRevB.71.035105}, 
originally developed to restore the Mott-Hubbard 
effects absent in the LDA description of
transition-metal  oxides.
A simplified formulation~\cite{PhysRevB.57.1505,PhysRevB.58.1201,
PhysRevB.71.035105}, 
in which the required Hubbard $U$ 
parameter is a  linear-response property of the system under 
scrutiny~\cite{PhysRevB.71.035105},
is now routinely and diversely applied~\cite{doi:10.1021/jz3004188,doi:10.1021/jp3107809,QUA:QUA24521,PhysRevLett.113.086402,PhysRevB.93.085135}.
\edit{Beginning with 
Ref.~\onlinecite{PhysRevLett.97.103001},
Marzari, Kulik, and co-workers have suggested and 
extensively developed~\cite{:/content/aip/journal/jcp/133/11/10.1063/1.3489110,
:/content/aip/journal/jcp/145/5/10.1063/1.4959882}
the interpretation of DFT+$U$ as a correction for SIE,
for systems in which it may
be primarily attributed to distinct subspaces 
(otherwise, the related Koopman's compliant 
functionals are available~\cite{PhysRevB.82.115121,
PhysRevB.90.075135}).
} %
The SIE correcting DFT+$U$ functional is given, where
$\hat{n}^{I \sigma} = \hat{P}^I \hat{\rho}^\sigma \hat{P}^I$, by
\begin{equation}
E_U  = \sum_{I \sigma} \frac{U^I}{2}
\text{Tr}
\left[ \hat{n}^{I \sigma}
- \hat{n}^{I \sigma } \hat{n}^{I \sigma } 
\right].
\label{Eq:dft+u}
\end{equation}
Here, $\hat{\rho}^\sigma$ is the Kohn-Sham density-matrix
for spin $\sigma$ and $\hat{P}^I$ is a projection operator for
the subspace $I$.
DFT+$U$ attains the status of an automatable, first-principles
method when it is provided with calculated 
Hubbard $U$ parameters~\cite{0953-8984-9-4-002,
PhysRevB.58.1201,PhysRevB.71.035105,
PhysRevB.74.235113,QUA:QUA24521}
(particularly at their self-consistency~\cite{PhysRevB.93.085135,
PhysRevLett.97.103001,doi:10.1021/jp070549l}),
which may be thought of as subspace-averaged
SIEs quantified in situ.
The subspaces are usually 
pre-defined for corrective treatment,
having been deemed responsible 
for the dominant SIEs on the basis of physical intuition
and experience, although
a further level of self-consistency over subspaces 
is possible using Wannier 
functions~\cite{PhysRevB.82.081102}.
DFT+$U$ effectively adds a set of penalty functionals
promoting integer eigenvalues in 
$ \hat{n}^{I \sigma}$, and it replicates the effect of
a derivative discontinuity in the energy, for each subspace $I$, by
adding an occupancy-dependent potential
$\hat{v}^{I \sigma} = U^I ( \hat{P}^I / 2 - 
\hat{n}^{I \sigma} )$.
%
%
%

\edit{While  DFT+$U$ 
is effective and computationally efficient, even 
linear-scaling~\cite{PhysRevB.85.085107},
a considerable degree of care is needed to calculate
the required $U$ parameters, which sometimes
pose numerical 
challenges~\cite{:/content/aip/journal/jcp/133/11/10.1063/1.3489110,QUA:QUA24521}.
Fully self-contained calculations 
of the Hubbard $U$ by means of 
automated variational extremization would be
extremely useful for many practitioners,   
and expedient in high-throughput materials search
contexts~\cite{curtarolo}.
The constraint-like functional form of DFT+$U$,
where the $U^I$ resemble the Lagrange multipliers
of penalty functionals on the eigenvalues of $\hat{n}^{I \sigma} $, 
suggests the possible viability of such a method.
Constrained density-functional theory 
(cDFT)~\cite{PhysRevLett.53.2512,Sit2007107,doi10.1021/cr200148b} 
formalizes 
and automates the use of self-consistent~\cite{PhysRevB.94.035159}
penalty functionals in DFT,
enforcing them as exact constraints by locating 
the lowest-energy compatible excited state of
the underlying  functional.
 %
%
%
It is  effective for treating SIE,  in its own right,
for systems comprising well-separated fragments, 
where it may be used to break physical symmetries 
and to explore the integer-occupancy states 
at which SIE is typically 
reduced~\cite{PhysRevLett.97.028303,doi10.1021/cr200148b}.
%
%
%
However, as we now demonstrate, 
cDFT is fundamentally incompatible with 
constraints beyond linear order, and therefore 
exact constraints cannot  be used  to 
correct SIE in an automated fashion.
As a result, it seems that we cannot excite a SIE affected system
to a state that will reliably exhibit less SIE, 
 without breaking a  symmetry.}

\edit{The simplest conceivable SIE-targeting} constraint 
functional  is the quadratic  form
$C_2 =\sum_I ( N^I - N^I_\textit{c} )^2$, where
\edit{$N^I = \textrm{Tr} [ \hat{n}^{I} ] $}
is the total occupancy of a particularly error-prone subspace $I$  
and $N^I_\textit{c}$ is its \edit{targeted} value, 
neglecting the spin index for concision.
\edit{This} constraint is a functional of  
\edit{ subspace total occupancies}, rather than the 
occupancy eigenvalues \edit{as in DFT+$U$},
\edit{which is an important  distinction for all but
single-orbital sites}.
For \edit{$N_\textrm{sites}$ symmetry-equivalent} subspaces with  
$N^I_\textit{c} = N_\textit{c}$ for all $I$,
the total-energy of the system \edit{is given by} 
  $W = E_\textrm{DFT} + V_\textit{c} C_2 $,
where $V_\textit{c}$ is a common cDFT Lagrange multiplier.
This gives rise to a constraining potential of the form
$\hat{v}_\textit{c} = 
2 V_\textit{c} \sum_I ( N^I - N_c ) \hat{P}^{I } $, 
making explicit \edit{its} dependence on the constraint non-satisfaction.
This, in turn, implies \edit{an}
externally \edit{imposed}  interaction correction given by
$\hat{f}_\textit{c}=
2 V_\textit{c} \sum_I \hat{P}^{I }  \hat{P}^{I} $,
which acts to modify the energy-density
profile, and 
which is identical to that generated by DFT+$U$
(Eq.~\ref{Eq:dft+u}) when
$V_\textit{c} = - U^I / 2$.
%

\begin{figure}
\includegraphics[width=1\columnwidth]{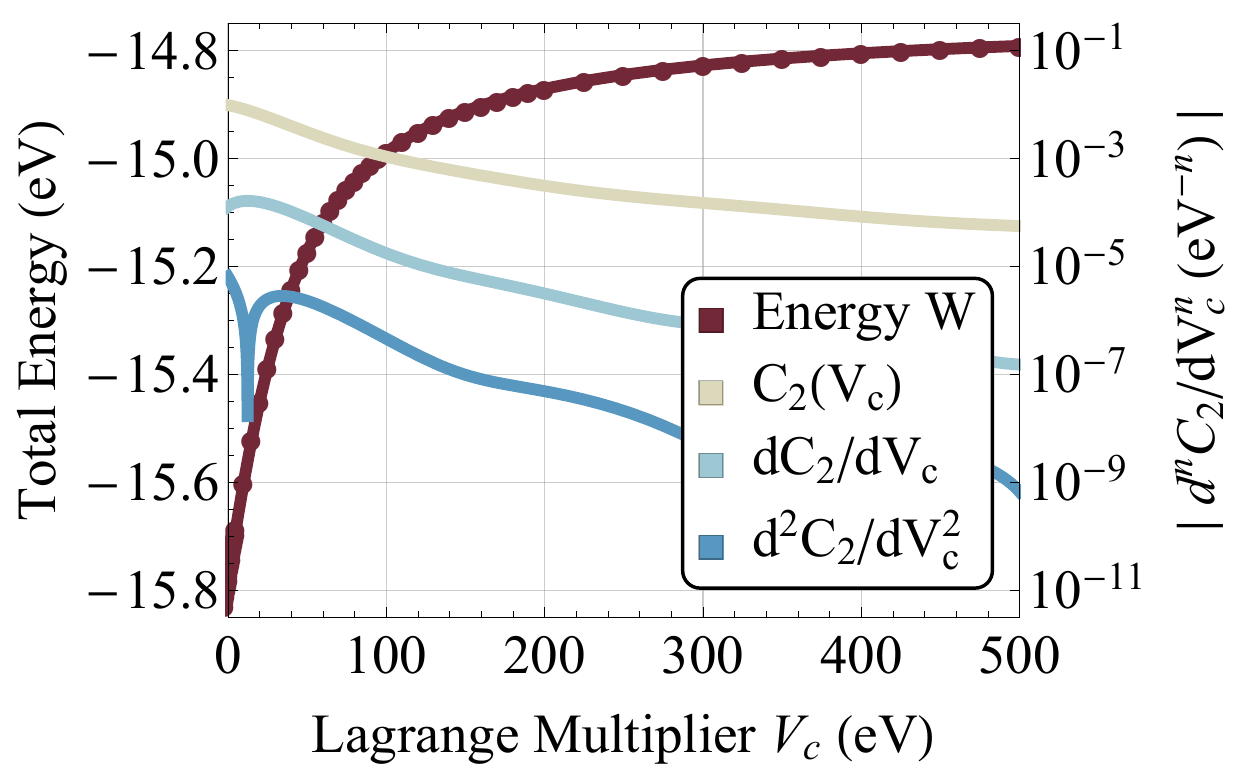}
\caption{(Color online) The constrained 
total-energy of the stretched $H_2^+$ 
system, with a target occupancy of $N_c=0.5$~e 
per fixed  atom-centered $1s$-orbital subspace,
against the cDFT Lagrange multiplier $V_c$. 
Also shown is the constraint functional $C_2=(N-N_\textit{c})^2$,
averaged over the two atoms, and its
first and second derivatives, which all  fall off rapidly with $V_c$.}
\label{figure1}
\end{figure}
%

Following Ref.~\onlinecite{PhysRevB.94.035159} 
for the self-consistent cDFT problem,
the Hellmann-Feynman
theorem provides that the first energy derivative is simply the 
constraint functional, i.e., 
$ d W / d V_{\textit{c}} = C_n $, so that the total-energy 
$W \left( V_\textit{c}  \right)$ always attains
a stationary point upon constraint satisfaction, 
in this case when $C_2 = 0$.
\edit{Fig.~\ref{figure1} depicts  this function for an ideal
 system for the study} of one-electron 
SIE~\cite{Cohen792,doi:10.1021/cr200107z}, $H_2^+$, 
simulated~\footnote{The DFT+$U$ 
functionality~\cite{PhysRevB.85.085107}
available in the \textsc{ONETEP} linear-scaling
DFT package~\cite{:/content/aip/journal/jcp/122/8/10.1063/1.1839852} 
was used with a hard ($0.65$~a$_0$ cutoff) 
norm-conserving pseudopotential~\cite{PhysRevB.41.1227}, 
$10$~a$_0$ Wannier function cutoff radii, and open 
boundary conditions~\cite{:/content/aip/journal/jcp/110/6/10.1063/1.477923}.} 
using the PBE functional~\cite{PhysRevLett.77.3865}. 
\edit{At the considered, intermediate bond-length of 4~a$_0$,  the
overlap of the two atom-centered PBE $1s$ orbital subspaces
yields a total occupancy double-counting of 24\%, 
accounting for spillage.}
\edit{The observed asymptotic behavior of $W \left( V_\textit{c} \right)$
demonstrates that the 
 $C_2$ constraint is unenforceable.}
\edit{Here}, a target occupancy of
$N_\textit{c} = 0.5$~e \edit{has been applied, 
necessitating} a repulsive constraint and a positive $V_\textit{c}$, but
the same qualitative outcome \edit{arises
for any $N_\textit{c} \ne N_\textrm{DFT}$.}

\begin{figure}
\includegraphics[width=1\columnwidth]{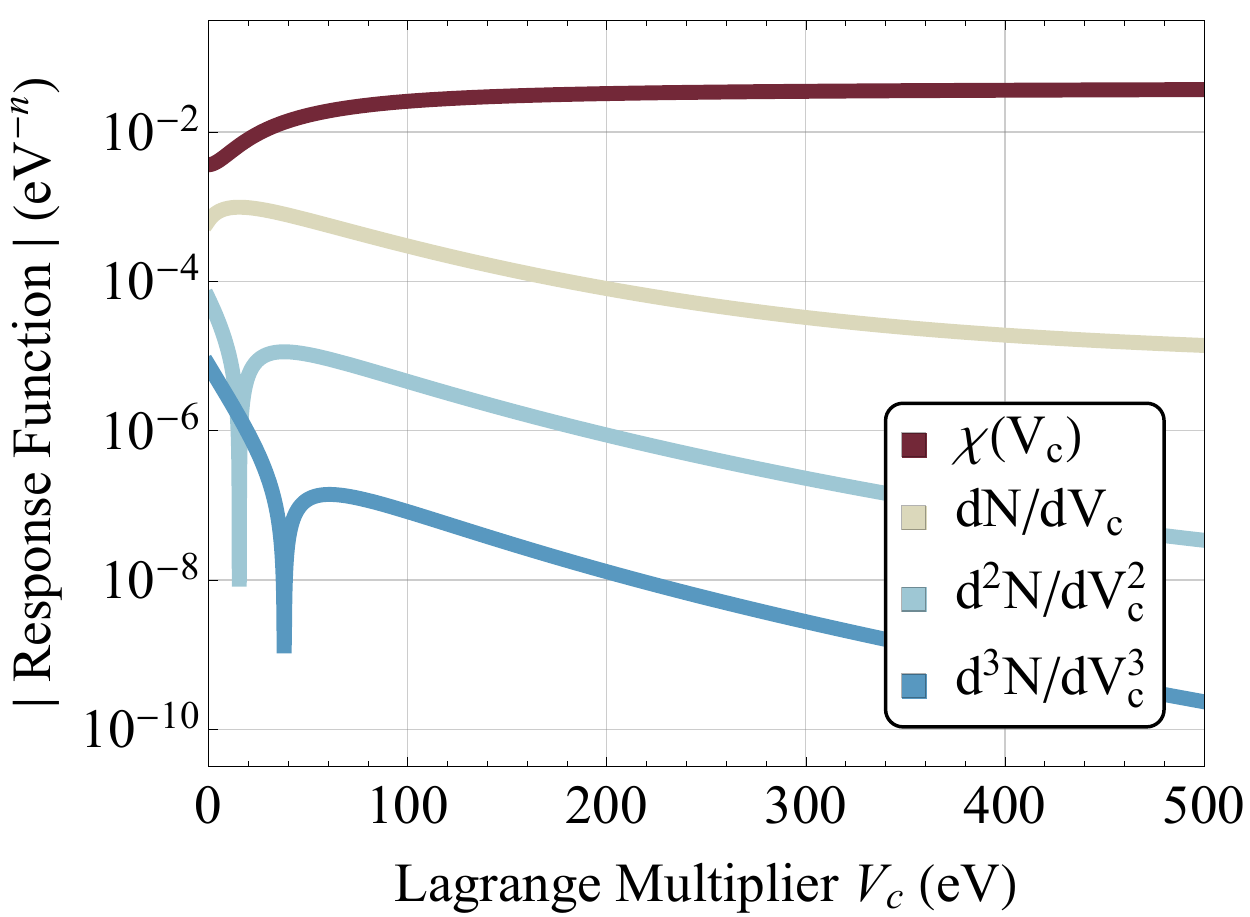}
\caption{(Color online) 
\edit{The magnitudes of the  
interacting density response $\chi$ 
and  cDFT response functions $d^m N / d V_\textit{c}^m$, 
calculated from a polynomial fit to the
average subspace occupancy for the same system 
as in Fig.~\ref{figure1}}. 
The $d^m N / d V_\textit{c}^m$
fall off  as we asymptotically approach constraint satisfaction,
while the occupancy, and hence $\chi$, tends to a constant value.}
\label{figure2}
\end{figure}

%
%
%
The key to the failure of the $C_2$ constraint is the 
\edit{fall-off of
all}  self-consistent cDFT response 
functions~\cite{PhysRevB.94.035159}
$d^n N / d V_\textit{c}^n = 
d^{n+1} W / d V_\textit{c}^{n+1}$, \edit{as depicted in Fig.~\ref{figure2}.
This results in a diminishing returns}
as $V_\textit{c}$ is increased, i.e.,  as the constraint 
asymptotically approaches satisfaction.
Motivated by \edit{this,} 
we investigate whether non-linear constraints of  form
$C_n = ( N - N_\textit{c} )^n$
are  unsatisfiable for any
order $n\neq1$ and for any
target choice $N_\textit{c} \ne N_\textrm{DFT}$.
The $n=0$ case is trivial, and the constraint 
is ill-defined for $n<0$
since there the total-energy diverges upon
constraint satisfaction.
$C_n$ becomes imaginary for non-integer $n$
with negative $( N - N_\textit{c})$, so that we may limit
our discussion to integers $n \ge 2$.
We begin by analyzing the derivatives of the total-energy $W
\left( V_\textit{c} \right)$.
The second derivative  follows directly from the above discussion, as
\begin{align}
\frac{d^2W}{dV_\textit{c}^2}= \frac{d C_n}{dV_\textit{c}} ={}& 
 n\left(N-N_\textit{c}\right)^{n-1}\frac{dN}{dV_\textit{c}}. %
\label{second_deriv}
\end{align}
The energy derivative of order $m$  generally involves cDFT
response functions up to order $m-1$, and
positive integer powers of $( N - N_\textit{c})$ 
which may vanish, depending on $m$ and $n$, but not diverge.
\edit{
The cDFT response function $dN / d V_\textit{c}$
may be gainfully expanded, if
$\hat{v}_\textit{ext}$ is the external potential,
in terms of the intrinsic
subspace-projected interacting response function
defined by $\chi = \mathrm{Tr} [ 
( d N / d \hat{v}_\textit{ext} ) \hat{P} ]$, 
since this object is independent of the form of the constraint.
The first-order cDFT response  $d N / d V_\textit{c}$
is thus expressed, 
by means of the chain rule in 
$ \hat{v}_\textit{ext} = \hat{v}_\textit{c} = 
V_\textit{c}(\delta C_n/\delta\hat\rho)
= 
 n  V_\textit{c} \left( N - N_\textit{c} \right)^{n-1}
\hat{P}
$, as}\edit{
\begin{align}
\frac{d N}{d V_\textit{c}}={}&
\mathrm{Tr} \left[ \frac{d N}{d \hat{v}_\textit{ext} }
\frac{d \hat{v}_\textit{ext} }{d V_\textit{c} } \right] =
n \chi \frac{d}{d V_\textit{c}} \left[ V_\textit{c} 
C_{n-1}\right] \nonumber \\
\Rightarrow \quad
\frac{d N}{d V_\textit{c}}={}&
n \chi C_{n-1}
\Big( 1 - n \left( n-1 \right) \chi
V_\textit{c} C_{n-2} \Big)^{-1},
\label{first_order_resp}
\end{align} 
an expression which we have verified numerically.
At any valid stationary point, $C_n = 0$ and 
each of $V_c$, $\chi$, and its derivatives must remain
finite.}
Thus, for $n \ge 2$,   the response $d N / d V_\textit{c}$ and 
energy curvature \edit{$d^2 W / d V_\textit{c}^2$ both then} 
vanish. The \edit{latter} is therefore not 
a stationary point discriminant, and we move to higher derivatives,
\edit{such as 
\begin{align}
\frac{d^3 W }{ d V_\textit{c}^3 } ={}&
n  \left(n-1\right) C_{n - 2} 
\left( \frac{d N}{d V_\textit{c} }  \right)^2 +
 n C_{ n - 1} \frac{d^2 N}{d V_\textit{c}^2 },
\end{align}
where the required second-order response is given by 
\begin{align}
\frac{d^2N}{dV_\textit{c}^2}={}& 
n \left[ \frac{d \chi}{d V_\textit{c}} C_{n-1}
+  \left( n -1 \right) \frac{d N}{d V_\textit{c}}  \right.
\nonumber \\  
{}& \times 
\left( \frac{d \chi}{d V_\textit{c}} 
 V_\textit{c} C_{n-2} +
2 \chi C_{n-2}  \right. 
 \nonumber \\  
{}& \quad + \left. \left.
 \left( n - 2 \right)  \chi 
 V_\textit{c} C_{\max \left ( n-3 , 0 \right)} \frac{d N}{d V_\textit{c}} 
\right)
  \right] \nonumber \\
  {}&\quad \quad \times
 \Big( 1 - n \left( n-1 \right) \chi V_\textit{c} C_{n-2} \Big)^{-1}.
\label{second_order_resp}
\end{align} 
\edit{This object, and thus $d^3 W / d V_\textit{c}^3 $}  both
vanish at stationary points for all $n \ge 2$, 
due to the vanishing $C_{n-1}$
in the first term of Eq.~\ref{second_order_resp}, 
 and due to the vanishing first-order 
 response (for which, see Eq.~\ref{first_order_resp})
 in all remaining terms.}

In general, the cDFT response
function $d^m N / d V_\textit{c}^m$ 
comprises terms proportional to response functions of the same type
but of lower order, 
plus a single term which is 
proportional to 
a \edit{potentially} non-vanishing mixed response 
function \edit{$d^{m-1} \chi / d V_\textit{c}^{m-1}$ multiplied by
the necessarily vanishing $C_{n-1}$.}
This serves as an inductive proof that response functions at
all orders, beginning with $dN/dV_\textit{c}$,
vanish as we approach a vanishing $C_n$, as
illustrated in Fig.~\ref{figure2}.
Then, since each term in the $m^{\textrm{th}}$ 
energy derivative is always
proportional to  non-divergent powers of $(N-N_\textit{c})$ and
response functions of at most order $m-1$, 
all energy derivatives tend to zero, 
as depicted in Fig.~\ref{figure1}, 
proving the conjecture.
\edit{Thus, non-linear constraints of SIE-targeting
$C_n$ form cannot be enforced}.

\edit{In order to cast the SIE-targeting $C_n$ 
functional into a viable form, one 
possible option remains.
We may 
expand} the single-site $C_2$, for example, as
$C_2=-2N_\textit{c}V_\textit{c}(N-N_\textit{c})-
V_\textit{c}(N_\textit{c}^2-N^2)$, 
\edit{and afford an additional degree of
freedom to the system by decoupling these two terms.}
Writing the result in the notation of DFT+$U$, by 
 change of variables, 
we arrive at \edit{the constraint energy
\begin{align}
 \sum_{I} \frac{U_1}{2}
\left(N^I-N_\textit{c}\right)
+ \sum_{I} \frac{U_2}{2}
\left(N^{ 2}_\textit{c}- N^{I 2}\right).
\label{Eq:dft+nu}
\end{align}
The vanishing response problem
is now circumvented, by interpreting the
Hubbard $U$ parameters
for linear and quadratic  terms 
as separate Lagrange multipliers. 
Adapting Eq.~\ref{Eq:dft+nu} to multiple, multi-orbital sites
and neglecting inter-eigenvalue terms, in the spirit of DFT+$U$, 
while retaining only the free-energy~\cite{PhysRevA.72.024502} 
 (setting $N^I_\textit{c}=0$), 
we arrive at the generalized DFT+$U$ correction given by
\begin{align}
E_{U_1 U_2} = \sum_{I \sigma} \frac{U^I_1}{2} 
\mathrm{Tr} \left[ \hat{n}^{I \sigma} \right]  
-  \sum_{I \sigma}  \frac{U^I_2}{2}
  \mathrm{Tr} 
\left[ \hat{n}^{I \sigma 2} \right].
\label{Eq:generalized}
\end{align}
Here, the DFT+$U$ functional of 
Eq.~\ref{Eq:dft+u} is recovered
by setting $U^I_1=U^I_2$.}
Otherwise, the corrective potential is modified to 
\edit{$\hat v^{I}_{U_1 U_2} = U^{I}_1\hat P/2-U^{I}_2\hat n^{I}$},
so that the characteristic occupancy eigenvalue
dividing an attractive from a repulsive  potential 
is changed from $1/2$ to $U^I_1/ 2 U^I_2 $. 
\edit{Self-consistency
effects aside, the $U_2$ 
parameters are responsible for correcting the interaction  
and for any gap modification, while the 
$U_1$ parameters may be used to adjust 
the linear dependence of the energy
on the subspace occupancies, and thereby to refine eigenvalue derived
properties such as the ionization potential.
We note a resemblance between Eq.~\ref{Eq:generalized} 
and the three-parameter DFT+$U\alpha\beta$ functional
proposed in Ref.~\onlinecite{dabo2008towards},  
where here a third degree of freedom may be 
retained by using $N^I_\textit{c} \ne 0$.}

\begin{figure}
\includegraphics[width=1\columnwidth]{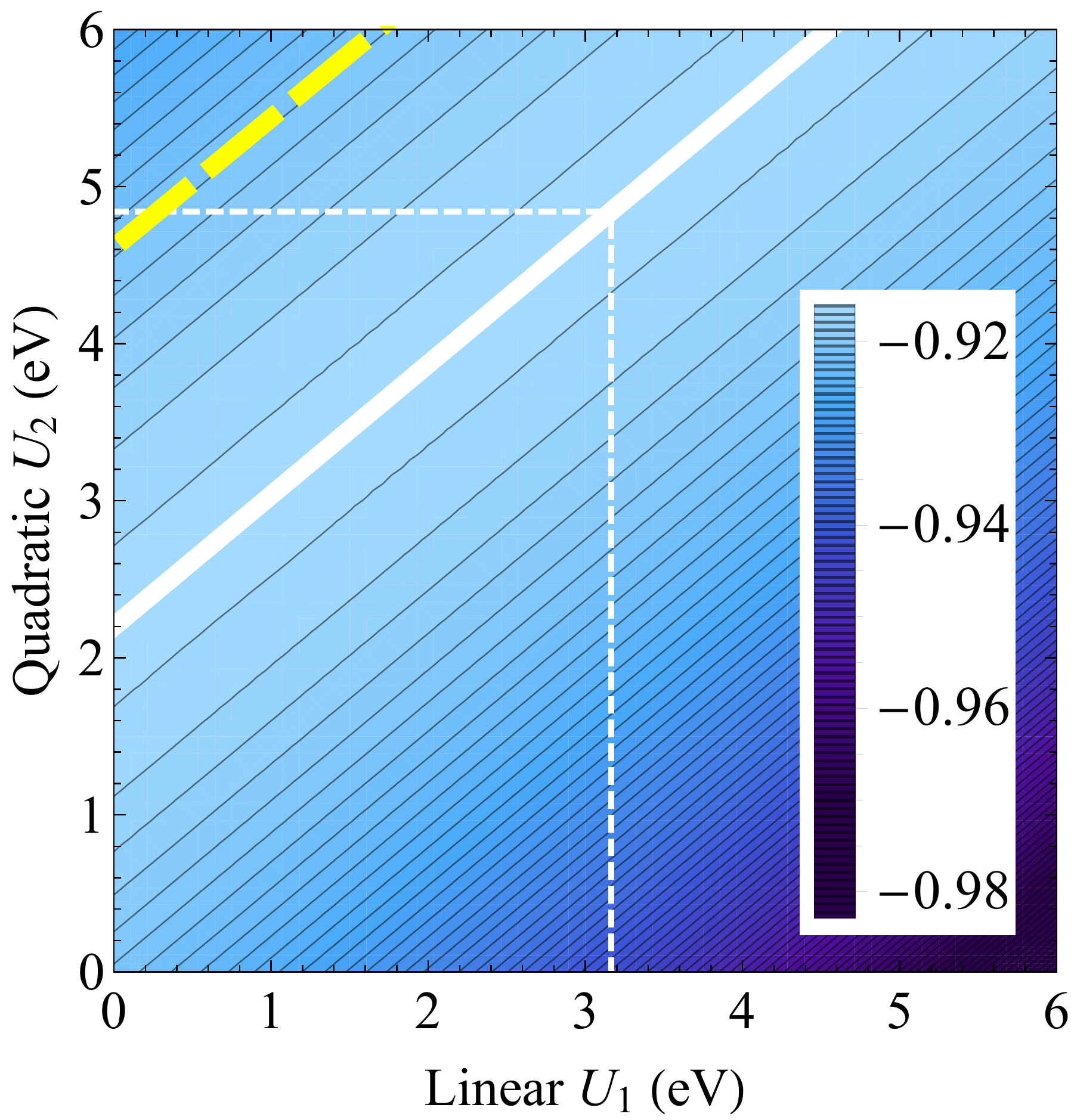}
\caption{(Color online) The constrained total-energy of
stretched $H_2^+$ against the Lagrange multipliers $U_1$
and $U_2$  defined in Eq.~\ref{Eq:dft+nu}. 
The subspace 
target occupancy is set to  $N_\textit{c} = N_\textrm{exact}$,
\edit{and the zero is set to the exact total-energy.}
The constraint is satisfied at the  
total-energy maximum along the 
solid white line.
\edit{The ionization potential is exact  
along the thick dashed line.} 
The linear-response Hubbard $U_2$, together with
the  $U_1$ value needed to correspondingly recover the exact 
subspace density, are shown using thin dashed lines.}
\label{figure3}
\end{figure}

Fig.~\ref{figure3}  shows the  total-energy $W$
of the $H_2^+$ system as before, but now against the $U_1$ and $U_2$
defined in Eq.~\ref{Eq:dft+nu}, with a
subspace target occupancy of 
$N_\textit{c}=N_\text{exact}=0.602$~e,
\edit{(the population of each of the
two PBE $1s$ orbital subspaces, calculated using the
exact functional).}
\edit{The total-energy is non-uniquely
maximized along the heavy white line
where the constraint is satisfied,
at  $\sim 0.92$~eV below the exact energy.}
 \edit{To understand why the total-energy is always degenerate, 
and hence why the occupancy condition under-defines the 
pair $\left( U_1,U_2 \right)$,}
it suffices to show that 
the Hessian of the constraint functional, 
 $H_{ij}=d^2 W/d U_i d U_j$~\footnote{Total-derivatives are used here
 to indicate that self-consistent
 density response effects are included. The Hubbard parameters
 remain independent variables.},
is  \edit{everywhere} singular. 
\edit{The determinant of $H_{ij}$ is
conveniently calculated
 in terms of the response functions, i.e.,
  by using the ground-state expressions
  $d W / d U_1 = N / 2$ and $d W / d U_2 = - N^2 / 2$, as
\begin{align}
|\bf{H}|
&=\frac{1}{2}\left|\begin{array}{cc}
	dN/dU_1 & -dN^2/dU_1 \\
	dN/dU_2 & -dN^2/dU_2 \\
	\end{array}\right| = 0 , 
\end{align}
as required, for all $U_1$ and $U_2$. This implies a vanishing 
energy curvature along the lines on which the corrective
potential is constant.
The linear-response Hubbard $U$ at this bond length, 
calculated using a method adapted from 
Ref.~\onlinecite{PhysRevB.71.035105},
is $4.84$~eV. If we intuitively set $U_2 = U$, then a corresponding
$U_1 = 3.16$~eV is required  to recover the exact subspace 
density. 
The line on which the ionization potential is exact
(for the special case of $H_2^+$, this means that the  
occupied Kohn-Sham state eigenvalue 
and the ion-ion energy 
sum to the exact total-energy, written $\varepsilon_{\textrm{DFT}}
+ E_\textrm{ion-ion} = E_\textrm{exact}$), 
intercepts $U_1 \approx 0$~eV 
at the linear-response $U_2 = U$, echoing the `SIC' 
double-counting correction proposed in Ref.~\onlinecite{PhysRevB.76.033102}.
Finally, for the constrained total-energy, 
we note that while it can
be tuned to reach a maximum at the exact energy for a plausible 
target,  $N_\textit{c} = 0.511$~e, an unreasonable 
$U_1= U_2 = -444.5$~eV is required to do so.
We   conclude, therefore, that 
an SIE affected ground-state cannot be systematically
excited to a state that is less so by means of exact constraints, 
without breaking a physical symmetry.
Put another way, the total-energy cannot typically be
SIE-corrected by altering the density alone, and a non-vanishing 
energy correction term is required.}

\begin{figure}
\includegraphics[width=1\columnwidth]{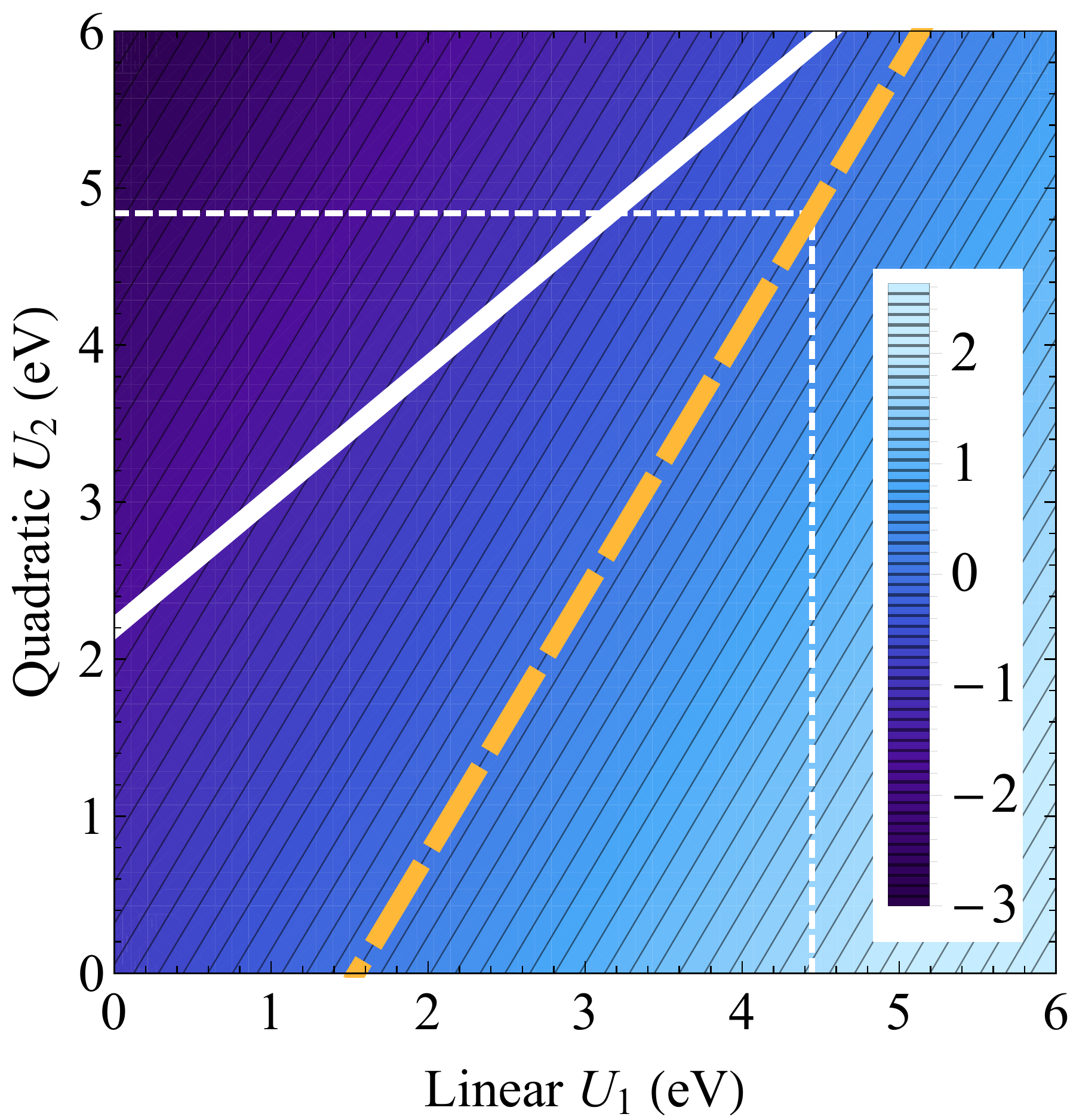}
\caption{(Color online) As per 
\edit{Fig.~\ref{figure3}}
 but showing the free-energy obtained 
by setting  $N_\textit{c} = 0$~e in Eq.~\ref{Eq:dft+nu}, \edit{i.e.,
using the generalized DFT+$U$  of Eq.~\ref{Eq:generalized}. 
The thick dashed
line is the exact energy intercept, and
the thin dashed lines show the linear-response $U_2$ together with
the corresponding $U_1$  needed to recover the exact energy.
The solid white line, as in Fig.~\ref{figure3},
indicates where the exact subspace
occupancy is recovered.}}
\label{figure4}
\end{figure}

\edit{Such a correction  is provided by  the
 generalized DFT+$U$ term  of Eq.~\ref{Eq:generalized}, 
 the total-energy generated by which is shown in Fig.~\ref{figure4}.
The zero of energy and the heavy dashed
line show $E_\textrm{exact}$, and
the thin dashed lines indicate the 
Hubbard $U_2 = 4.84$~eV and corresponding
$U_1 = 4.44$~eV required to recover it.
$E_\textrm{exact}$ is attained by a traditional DFT+$U$ calculation
 at  $U_1=U_2 = 3.85$~eV, and 
the intersection of the heavy
solid and dashed lines yields  the pair, 
$U_1=5.73$~eV and  $U_2=6.98$~eV,
at which $E_\textrm{exact}$ and $N_\textrm{exact}$
are located.
At the same point, the Kohn-Sham eigenvalue 
$\varepsilon_\textrm{DFT}$
lies  at $\sim 2.8$~eV above $\varepsilon_\textrm{exact}$,
reflecting that an accurate total-energy at a particular
occupancy may coincide with an
inaccurate ionization energy, and vice versa,
as was recently shown 
in detailed analyses of the residual SIE in hybrid 
functionals~\cite{PhysRevB.94.035140}
and in DFT+$U$ itself, at fractional  total  
occupancies~\cite{:/content/aip/journal/jcp/145/5/10.1063/1.4959882}.}
%

\begin{figure}
\includegraphics[width=1\columnwidth]{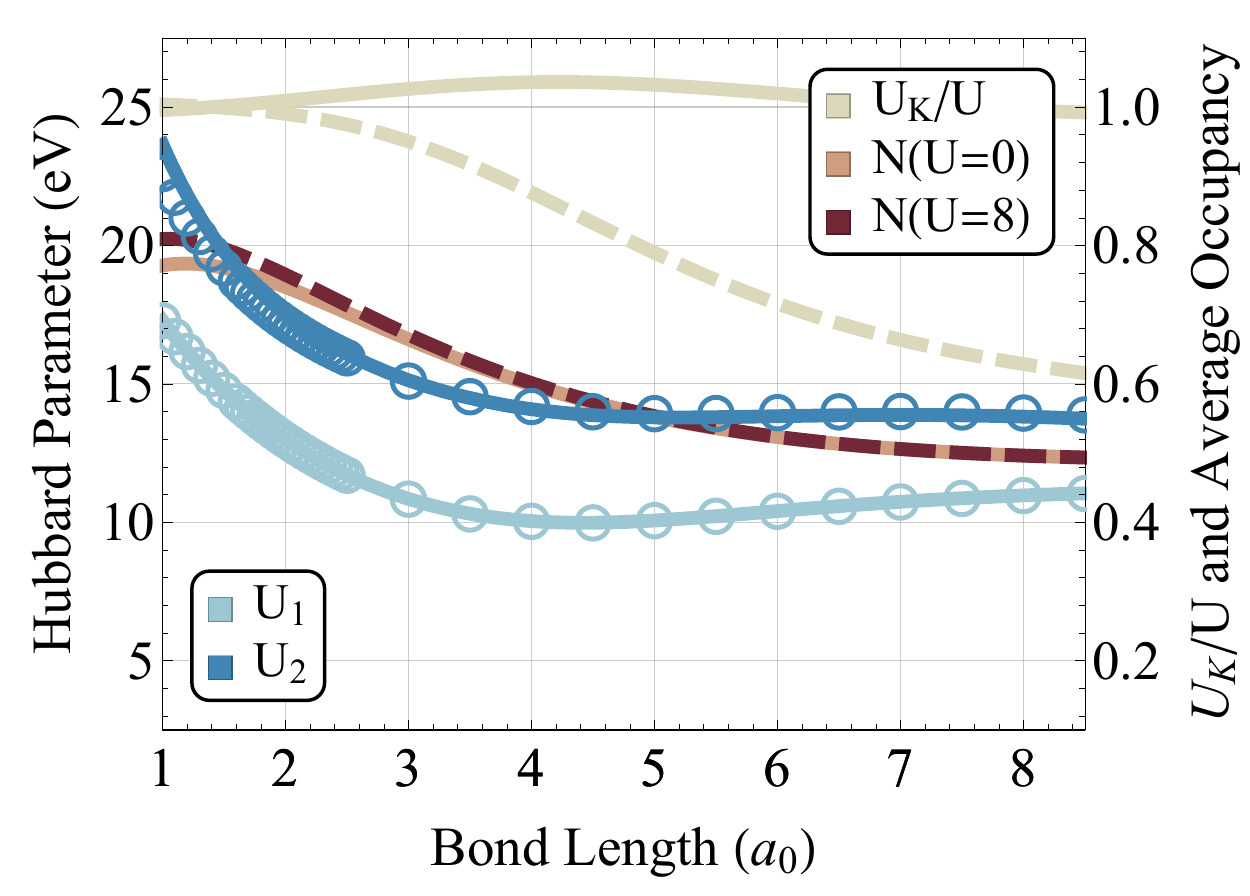}
\caption{\edit{
(Color online) Generalized Hubbard $U$
parameters estimated for the PBE $H_2^+$ molecule at varying
bond lengths.
Approximately parallel solid curves (blue, left) 
depict the $U_1$ and $U_2$ values required to recover 
the exact total-energy and Koopmans' condition, 
assuming constant PBE subspace occupancies.
Open circles show the corresponding quantities calculated using a 
practical scheme based on the conventional Hubbard $U$ and the PBE occupied eigenvalue (see text). 
The topmost curves (sand, right) show the fraction of the latter
$U_1$ (dashed) and $U_2$ (solid) due to the Koopmans 
term, $U_K$. 
On the same axis (dark red, right), we show that the average 
DFT+$U$ subspace occupancy is insensitive to the $U$ value.}
}
\label{figure5}
\end{figure}

%
%
%

\edit{The generalized DFT+$U$ functional  enables 
simultaneous correction of the ionization potential and  
total-energy, or the correction of either together with 
Koopmans' condition~\cite{PhysRevB.23.5048,PhysRevLett.49.1691,
PhysRevB.82.115121}.
In the one-electron case, as in $H_2^+$, 
Koopmans' condition may be enforced at $\varepsilon_\textrm{exact}$.
The approximately parallel solid curves of
Fig.~\ref{figure5} illustrate the large $U_1$
and $U_2$ values required to do so, 
as a function of internuclear distance.
To estimate these, we have  used 
the convenient feature of $H_2^+$ 
that the PBE $1s$ orbital subspace projectors closely match
Kohn-Sham orbitals in spatial profile, 
almost exactly so at dissociation.
As demonstrated by the  
occupancy curves in Fig.~\ref{figure5}, 
this implies a negligible charge and kinetic self-consistency effect,
vanishing entirely for $U_1  =  
2 N_\textrm{DFT} U_2 $,
in subspace-uniform corrections such as those in question.
%
For $N_\textrm{sites}$ equivalent one-orbital 
subspaces spanning the
energy window responsible for 
 $\varepsilon_{\text{DFT}} $, 
the density non-self-consistent 
$U_1$ and $U_2$ are  derived from  
$E_{\text{exact}} \approx E_{\text{DFT}} + N_{\textrm{sites}}
\left( U_1 N_{\text{DFT}} - U_2 N_{\text{DFT}}^2 \right) / 2 $, 
where $N_\textrm{DFT} = \mathrm{Tr} 
\left[ \hat{n}_\textrm{DFT} \right]$, 
and  
$\varepsilon_{\textrm{exact}} \approx \varepsilon_{\text{DFT}} + 
 \left( U_1 - 2 U_2 N_{\text{DFT}} \right) / 2$,
in which  the subspace overlap and spillage are also neglected.}

\edit{Since subspace response stiffening very typically 
results from the application of a
 conventional DFT+$U$ correction~\cite{doi:10.1021/jp070549l,
 PhysRevLett.97.103001}, it is
promising to construct a charge non-self-consistent first-principles
calculation scheme for  $U_1$ and $U_2$, e.g., for use in 
refining DFT+$U$ calculations in 
order to approximately enforce Koopmans' condition.
%
For this, let us suppose we have calculated 
a conventional $U$ which 
 reconciles the total energy reasonably, so that
$E_{\text{exact}} \approx E_{\text{DFT}} +  U N_{\text{sites}}\left( N_{\text{DFT}} - N^2_{\text{DFT}} \right) / 2 $.
We may combine this with the previous 
two equations and a further requirement for 
Koopmans' condition at an accurate energy, i.e., 
$\varepsilon_{\text{exact}} = 
E_\textrm{DFT} [N] - E_\textrm{DFT} [N-1]$, 
where the latter is the DFT-estimated 
total-energy of the ionized system.
This results  (see Fig.~\ref{figure5} open circles for data) in 
\begin{align}
U_1 &{}\approx  U \left( 1 - N_{\text{DFT}} \right) 
\left( 2 - N_\textrm{sites} N_{\text{DFT}}  \right) 
+ U_K, \;\; \mbox{and} \nonumber \\
N_{\text{DFT}} U_2 &{}\approx  U \left( 1 - N_{\text{DFT}} \right) 
\left( 1 - N_\textrm{sites} N_{\text{DFT}}  \right)
+ U_K , 
\label{Eq:U1U2}
\end{align}
where, with 
 $ U_K =  2 \left( E_\textrm{DFT}[N-1] -
 E_\textrm{DFT}[N] + 
\varepsilon_\textrm{DFT} \right) $,  we define  the
 `Koopmans $U$'.
We emphasize that  
only convenient, approximate DFT quantities are used in
these formulae.
The interdependence $U_1 - U_2 N_\textrm{DFT} 
\approx U \left( 1 - N_\textrm{DFT} \right)$,
for any value of $N_\textrm{sites}$, reveals the role
of $U$ in splitting $U_1$ and $U_2$.
In $H_2^+$, Koopmans' condition
pushes both up to considerably higher values
than are commonplace~\cite{QUA:QUA24521} 
for the conventional 
$U$ of DFT+$U$, which,
following 
Ref.~\onlinecite{:/content/aip/journal/jcp/145/5/10.1063/1.4959882}
and given $N_\textrm{PBE}$,
  lies close to its regime of minimal efficacy
for eigenvalue correction. 
%
%
The Koopmans fraction of each parameter, 
$U_K / U_1$ or $U_K / \left( N_\textrm{DFT} U_2 \right)$,
generically denoted by `$U_K  / U$' in Fig.~\ref{figure5}, lies close
to unity for $U_2$ at all $H_2^+$ bond lengths, 
and it exceeds unity slightly when the
$U_K$ and $U$-related 
contributions tend to cancel.
$U_1$ is also $U_K$-dominated at short 
bond lengths,
at which $U_1 \approx N_\textrm{DFT} U_2$, before
ultimately falling off to the average of $U$ and $U_2$ in
the fully dissociated limit.
%
If the outlined proposed scheme is applied to finesse
an existing DFT+$U$ calculation that
 is already 
accurate for recovering the total-energy,
using a linear-response~\cite{PhysRevB.71.035105,
PhysRevLett.97.103001} 
or otherwise calculated $U$, call it $U_0$, 
then $U = 0$~eV is the appropriate value to use
in our approximate formulae of 
Eq.~\ref{Eq:U1U2}.
An approximately Koopmans' compliant  DFT+$U$ calculation
then results from the use of the parameters $U_0 + U_K$ and $U_0 + 
 U_K / N_\textrm{DFT+$U_0$}$, 
in place of $U_1$ and $U_2$.
The proposed scheme may  be generalized to 
multi-orbital subspaces straightforwardly, in terms of the
eigenvalues of $\hat{n}^I_\textrm{DFT}$ instead of $N_\textrm{DFT}$.
%
The constant-$N_\textrm{DFT}$ approximation  may  
 be replaced by a linear-response
approximation, in terms of $\chi$,
or lifted entirely by means of
a  parametrization of the occupancies 
and a numerical solution of 
the resulting equations.}
%

\edit{To conclude, we have proven
analytically, with stringent numerical tests,
that non-linear constraints are incompatible with cDFT.
It is not possible, therefore, to automate 
systematic SIE corrections of DFT+$U$ type by means
of cDFT, notwithstanding the great utility of the latter, e.g.,
for correcting SIE by 
promoting broken symmetry, integer-occupancy configurations
well described by approximate 
functionals~\cite{PhysRevLett.97.028303,doi10.1021/cr200148b}.
%
Nonetheless, we have found that the cDFT free-energy
functionals, dubbed `generalized DFT+$U$' functionals, 
offer the intriguing capability of simultaneously 
correcting  two central quantities in DFT, the total-energy
and the highest occupied orbital energy.
Our approximate formulae for the required
parameters, which may differ greatly from the 
familiar Hubbard $U$, offer a framework within which to further
develop double-counting techniques and first-principles 
schemes for the promising class of SIE correcting 
methods based on DFT+$U$~\cite{QUA:QUA24521,
PhysRevLett.97.103001,
:/content/aip/journal/jcp/145/5/10.1063/1.4959882,
:/content/aip/journal/jcp/133/11/10.1063/1.3489110}, 
as well as opening up possibilities for their diverse application.
We envisage that SIE  correction schemes
of two or more parameters may also be useful 
for generalizing the exchange fraction of
hybrid functionals~\cite{PhysRevB.94.035140}, 
and for DFT+$U$ type corrections of perturbative many-body 
approximations such as $GW$~\cite{PhysRevB.82.045108}, 
the deviation from linearity of which is somewhat 
analogous to that of approximate DFT~\cite{PhysRevA.75.032505,PhysRevB.93.121115}.
For the analysis and correction of spuriously self-interacting
multi-reference systems,  we may
learn much from the exact solution of minimal  
models~\cite{:/content/aip/journal/jcp/140/16/10.1063/1.4871875}.}
%
%
%
%
%
%

This work was enabled by the Royal Irish Academy -- Royal Society 
International Exchange Cost Share Programme (IE131505). 
GT acknowledges support from EPSRC UK 
(EP/I004483/1 and EP/K013610/1). 
GM and DDO'R wish to acknowledge support from the Science Foundation 
Ireland (SFI) funded centre AMBER (SFI/12/RC/2278).
All calculations were performed on the Lonsdale cluster 
maintained by the Trinity Centre for High Performance Computing. 
This cluster was funded through grants from Science Foundation Ireland.

\end{document}